\documentclass[11pt]{article}

\newcommand{\mb}{\mathbf}
\newcommand{\bs}{\boldsymbol}

\usepackage{graphicx}

\usepackage{soul}
\usepackage{natbib}
\setcitestyle{authoryear,open={(},close={)}}

\usepackage{hyperref}
\usepackage{color}
\usepackage{multirow}

\usepackage{graphicx}
\usepackage{amsmath}
\usepackage{bbm}
\usepackage{comment}
\usepackage{setspace}
\newcommand{\blind}{1}

\begin{document}
\def\spacingset#1{\renewcommand{\baselinestretch}%
	{#1}\small\normalsize} \spacingset{1}

\if1\blind
{
	\title{\bf Assessing the Reliability of Wind Power Operations under a Changing Climate with a  Non-Gaussian Bias Correction}
	\author{Jiachen Zhang\\
	Department of Applied and Computational Mathematics and Statistics,\\
		University of Notre Dame (USA)\\
		and \\
		Paola Crippa\\
		Department of Civid and Environmental Engineering and Geoscience,\\
		University of Notre Dame (USA)\\
		and\\
		Marc G. Genton\\
		Statistics Program,\\
		King Abdullah University of Science and Technology (Saudi Arabia)\\
		and\\
		Stefano Castruccio\thanks{scastruc@nd.edu}\hspace{.2cm}\\
		Department of Applied and Computational Mathematics and Statistics,\\
		University of Notre Dame (USA)\\
		\\
	}
	\maketitle
} \fi

\newpage

\begin{abstract}
\spacingset{2}
\quad Facing increasing societal and economic pressure, many countries have established strategies to develop renewable energy portfolios, whose penetration in the market can alleviate the dependence on fossil fuels. In the case of wind, there is a fundamental question related to the resilience, and hence profitability, of future wind farms to a changing climate, given that current wind turbines have lifespans of up to thirty years. In this work we develop a new non-Gaussian method to adjust assimilated observational data to simulations and to estimate future wind, predicated on a trans-Gaussian transformation and a cluster-wise minimization of the Kullback–Leibler divergence. Future winds abundance will be determined for Saudi Arabia, a country with a recently established plan to develop a portfolio of up to 16 GW of wind energy. Further, we estimate the change in profits over future decades using additional high-resolution simulations, an improved method for vertical wind extrapolation and power curves from a collection of popular wind turbines. We find an overall increase in daily profit of \$272,000 for the wind energy market for the optimal locations for wind farming in the country. 
\end{abstract}

\noindent%
{\it Keywords:} Bias correction; Kullback-Leibler divergence; Non-Gaussian process; Nonstationary model; Spatio-temporal model; Wind energy
\vfill

\spacingset{2} 

\addtolength{\textheight}{.5in}%

\section{Introduction}

The vast evidence of the negative effects of fossil fuel emissions \citep{ipc14} calls for a major systemic change in current strategies to produce and distribute energy throughout the world. Societies worldwide are adapting by developing renewable alternatives to reduce dependence on fossil fuels and to align with the standards imposed by the Paris Agreement \citep{kin17}. Wind has been the natural resource with the largest share of power generation worldwide, with the United States and China being the two major leaders \citep{ren18}. In the United States, renewable energies, with wind energy being a major contributor, are predicted to surpass coal in terms of share of the energy market by the end of this decade. Similarly, while European countries have an overall smaller absolute installed capacity, the penetration of the energy in the national grid is in percentage considerable, with a peak of 42\% in Denmark. 

The wind energy sector is at its early stages in developing countries such as Saudi Arabia. Despite being one of the countries with the highest per capita energy consumption \citep{wb20}, the sixth largest consumer of oil worldwide \citep{bp20} and the largest in the Gulf Cooperating Council (GCC) \citep{ire18}, the transition to renewable energy is very recent and almost exclusively focused on solar energy \citep{ire19}. Currently Saudi Arabia's contribution to the GCC's renewable energy portfolio amounts to 16\% of the total capacity, with only 0.2\% available for sharing, and only 2\% generated from wind \citep{ire18}. As part of the recently outlined `Vision 2030' plan \citep{nur17}, Saudi Arabia aims to generate 16 GW of wind energy \citep{nre18}, positioning the country as one of the major global wind energy suppliers and considerably contributing to the planned emissions reduction targets stipulated in the United Nations Framework Convention on Climate Change \citep{unf20}. Investments in wind energy would also result in the increased reliability of renewable energy, as for example solar energy is only available during daytime, whereas winds often peak at nighttime.

Under the aforementioned scenario, a comprehensive analysis must be conducted to identify optimal sites for developing wind farms based on a cost-benefit analysis. Several recent studies with global climate models at annual \citep{jeo18} and monthly \citep{jeo19} scale, and later at daily level \citep{tag19,tag21}, including validation of observational and simulated data sets \citep{che18} and extreme wind conditions \citep{che20}, provided initial evidence about the availability of sufficient wind for installing wind turbines. Very recently, \cite{gia20} conducted the first full feasibility study with a new high-resolution ensemble and identified the optimal locations and types of turbine based on maintenance and operational costs. While their study provided the first detailed assessment of the country's resources, it relied on only four years of data, from 2013 to 2016, due to computational and storage constraints. This timescale is inconsistent with the current lifespan of wind turbines, which can be as long as thirty years, and the multidecadal effects of climate change could have an impact on the wind resource availability. 

In order to provide a pathway for a feasible and robust implementation of a wind energy portfolio, the question of resilience under a changing climate must be addressed. Given the computational impossibility to simulate high resolution numerical simulations for decades, an alternative strategy focused on publicly available data must be devised. In this work we focus on developing a methodology for assessing future winds, predicated upon the estimation of the relationship between simulated and observed data for a historical period at the same spatial and temporal scales. Under the assumption of an enduring relation between these two data sets, future wind behaviors are estimated by applying the same relation to future climate simulations. 

Methods for adjusting observations and simulations have a long history in geoscience (see, e.g., \cite{yua19, kim15} and \cite{ho12,haw13} for a general review), and bear some resemblance but also substantial differences with methodologies for calibrating observations with numerical simulations. Adjustment methods can be generally divided in two categories: observation-driven or simulation-driven. 

Observation-driven approaches focus on adjusting observations based on estimated changes in simulations. This \textit{Delta method} or \textit{change factor} \citep{haw13} has been subject to some recent generalizations beyond a simple mean correction. Indeed, \cite{lee15} proposed a spectral-based approach to the adjustment of time series, thus resulting in an adjustment of the implied covariance structure. At the core of this method is the notion of (penalized) estimation of the ratio between the spectral densities of observations and simulations. \cite{pop16} proposed a generalization of the aforementioned approach to a transient climate and multiple simulations under different future scenarios.

Simulation-driven methods are instead focused on estimating an empirical relationship between historical observation and simulation, and use it to correct future simulations. This \textit{bias correction} approach, in its simplest form, focuses on an individual time series, estimates the differences between observational and simulated historical climate, and applies this difference to future simulations to obtain the future observations. Slightly more sophisticated approaches involve also the use of variance (see Section \ref{sec:rev} for a comprehensive review), or transforming the quantiles of the distribution (\textit{quantile mapping}, \cite{can15}). In more recent years, more studies have acknowledged the need for methods of bias adjustment for spatially distributed (and possibly multivariate) data. \cite{ngu19} proposed a frequency based adjustment of spatial data. Among others, \cite{meh16,can18} and \cite{can20} extended quantile mapping to multivariate time series, and \cite{fra20} provided a recent overview and comparison of multivariate bias correction methods. 

In this work, we propose a new bias correction approach based on a non-Gaussian clusterwise spatial transformation, which is estimated by minimizing the distance between the joint distribution of the observational and of the simulated data field. Our proposed approach is based on three fundamental ideas: 1) non-Gaussianity can be accounted for through a marginal transformation which is slowly varying in space; 2) the transformation to Gaussianity can be inferred by minimizing the distributional distance between the two fields, as long as this value is comparable, and possibly within the confidence interval of the maximum likelihood estimate; and 3) transformation to Gaussianity allows a simple adjustment of first and second moments and a back-transformation to the original scale. 

\begin{sloppypar}
Section \ref{sec:prepr} introduces the data sets used for this study, validates the historical wind speeds across Saudi Arabia, and proposes a model for the mean and temporal dependence. Section \ref{sec:rev} reviews current approaches to observation-simulation corrections. Section \ref{sec:mod} introduces the proposed methodology for non-Gaussian adjustment. Section \ref{sec:valid} validates the model with a simulation study with non-Gaussian random fields and historical data. Section \ref{sec:app} applies the proposed methodology, along with vertical wind extrapolation and power curve evaluation, to estimate the change in daily profits from future winds in Saudi Arabia. Section \ref{sec:conc} concludes with a discussion. The code for this manuscript is available at the following GitHub repository: github.com/jiachenzhang001/Non-Gaussian-Bias-Correction.
\end{sloppypar}

\section{Data and preprocessing}\label{sec:prepr}

In this study, we focus on daily wind speed data at 10 meters above the ground level in Saudi Arabia, which is bounded approximately by longitudes of $34^{\circ}E–56^{\circ}E$ and latitudes of $16^{\circ}N–33^{\circ}N$ (see Figure S1(a-b)). Wind speed is derived as the Euclidean norm from the zonal and meridional velocity  (i.e., over the $x$ and $y$ axes). We present the data set of the power curves along with the information related to the cost of energy in the supplementary material.

\subsection{Observational data}
We use the Modern-Era Retrospective Analysis for Research and Applications, version 2 (MERRA-2, \cite{gel:17}), available from 1980 to the present day. Reanalysis data consists of observational data assimilated to a numerical weather model and, in the geoscience community, this data product is considered the best representation of the state of the Earth's system. MERRA-2 is the reference observational data set used in our study and is available on a regular grid with a resolution of $0.625^{\circ}\times 0.5^{\circ}$ in longitude and latitude, respectively. We only use daily wind speed data from 1980 to 2005, for a total of 26 years, to match the simulation data sets presented in Section \ref{sec:menacordex}. There are $n=614$ locations in Saudi Arabia at MERRA-2 resolution for a total of $T=9,497$ days. Throughout this manuscript, we will denote the MERRA-2 wind fields as $W_O(\mb{s}_i,t_j)$, where the subscript $O$ indicates the observation. 

\subsection{Regional simulations}\label{sec:menacordex}

We use the simulations from the Coordinated Regional Climate Downscaling Experiment (CORDEX), which is a set of coordinated regional experiments. Specifically, we focus on the Middle East North Africa (MENA) CORDEX Program, and among the five available simulations, we select CORDEX-4, which exhibits the best agreement with MERRA-2 in Saudi Arabia, as demonstrated by \cite{che18}. The simulation resolution is $0.22^{\circ}\times 0.22^{\circ}$, finer than MERRA-2, and boundary conditions for the historical and future periods are provided by the Geophysical Fluid Dynamics Laboratory global coupled climate-carbon Earth System Models (GFDL-ESM2M, \cite{dun13}). Future GFDL-ESM2M runs are simulated under the Representative Concentration Pathways 8.5 (RCP 8.5, \cite{van11}) scenario, in which the global radiative forcing is assumed to increase by 8.5 $W/m^2$ by 2100 \citep{cmip5}. We regrid and upscale the MENA CORDEX data to the MERRA-2 resolution by considering the average of locations in MENA CORDEX that are within the range of two consecutive MERRA-2 grids. Even though the historical run spans from 1950 to 2005, we only consider data from 1980 to 2005 to align the observation window with MERRA-2; for future runs we only consider simulations in the near future for the same number of years, i.e., from 2025 to 2050. We will denote the MENA-CORDEX wind fields as $W_S(\mb{s}_i,t_j)$, where the subscript $S$ indicates for simulation.

\subsection{Model for the mean and temporal dependence}\label{sec:meantemp}

We assume a periodic climatology described by $K$ harmonics in order to account for the inter-annual wind variability. If we denote the wind speed for location $\mb{s}$ and time $t$ by $W(\mb{s},t)$, its interannual variability across the year can be written as 
\begin{equation}\label{eq:trend}
\begin{array}{rcl}
  W(\mb{s}_i,t_j) & = & \mu(\mb{s}_i,t_j)+\sum_{i'=1}^P \phi_{i',i} \varepsilon(\mb{s}_i,t_{j-i'}) +\varepsilon(\mb{s}_i,t_j),\\[7pt]
  \mu(\mb{s}_i,t_j) & = & \omega_i* \text{yr}(t_j)+\sum_{k=1}^K \left\{\beta_{k,i} \sin\left(\frac{2\pi k t_j}{\delta}\right)+\beta'_{k,i} \cos\left(\frac{2\pi k t_j}{\delta}\right)\right\},
 \end{array}
\end{equation}
where $\delta\in \{365,366\}$ depending on whether the year is non-leap or leap, $\mb{s}_i=(x_i,y_i),  i=1,\ldots, n; j=1,\ldots, T$, and $\text{yr}(t)$ represents the year of day $t$. Thus, the model assumes a location-specific linear annual trend, with $K$ harmonics to explain inter-annual variability and $P$ autoregressive coefficients. We further assume that $\varepsilon(\mb{s}_i,t_j)$ is independent across $t_j$ with a spatial dependence that will be specified later. 
The mean and the autoregressive coefficients are estimated through site-specific inference, initially by assuming Gaussian errors and independence in time for estimating $\{\omega_i,\beta_{k,i},\beta'_{k,i}\}$ and subsequently estimating $\phi_i$ by maximum likelihood. In the supplementary material we present diagnostics to show that 1) the linear trend $\omega_i$ is significant for the majority of points in both data sets (Figure S3); 2) three harmonics ($K=3$) are sufficient to explain the climatology for both data sets (Figure S4); 3) the parameters $\beta_{k,i}, \beta'_{k,i}, \phi_i$ are constant in time (Figures S5, S6 and S7); and 4) the residuals of \eqref{eq:trend} are uncorrelated in time (Figure S8).

\section{Review of adjustment approaches}\label{sec:rev}

MERRA-2 represents the state of the system assimilated from observations, so it can be observed up to the present and its future needs to be estimated. The simulations obtained from MENA CORDEX can be used as a proxy to assess the future wind speed. From the preliminary analysis in supplementary material, a systematic mismatch between the two data sets is apparent. The objective of this study is to provide an adjustment of this mismatch in the historical period and use it to predict future MERRA-2 observations using the available future MENA CORDEX simulations. While the methods are general, for consistency with the previous section, we will denote the spatio-temporal wind fields as $\mb{W}^{(H)}_O$ and $\mb{W}^{(H)}_S$ for MERRA-2 (\textit{O} for observations) and MENA CORDEX (\textit{S} for simulations), respectively, for the historical period (1980-2005). The future data will be denoted as $\mb{W}^{(F)}_O$ and $\mb{W}^{(F)}_S$. Throughout this section, we will refer to MERRA-2 as `observations' and MENA CORDEX as `simulations' to emphasize the generality of our methodology. 

\subsection{Correction for marginal distributions}

The simplest approach to achieve adjusted future simulations is bias correction, which assumes a simple additive bias between observations and simulations, estimates it in the historical period, and then adjusts future simulations. Bias correction has been widely used in a number of studies in geoscience (see, e.g.,  \cite{hemer,lian} for wind applications). This approach assumes that there is no difference in marginal variance, spatial covariance, or other high-order moments. Formally, we assume that the observation-corrected data for a location $\mb{s}_i$ can be expressed as
\begin{equation} \label{bias:mean}
  W^{(F)}_O(\mb{s}_i,t_j)=W^{(F)}_S(\mb{s}_i,t_j)+\left\{\mu^{(H)}_O(\mb{s}_i,t_j)-\mu^{(H)}_S(\mb{s}_i,t_j)\right\},
\end{equation}
where $t_j$ refers to the $j$th time for the historical or future time period, and $\bs{\mu}^{(H)}_{\ell}(\mb{s}_i)$ for $\ell=\{O,S\}$ is the mean in the historical period (from 1980 to 2005), which is estimated via the location-specific inference of the mean parameters in \eqref{eq:trend}, as described in Section \ref{sec:prepr}.

Two data sets typically exhibit differences also in their temporal variability (see Figure S2), which cannot be accounted for by a simple bias correction. Therefore, a relatively more articulated approach focuses on adjusting both the mean and variance \citep{teu,li}. Formally:
\begin{equation} \label{bias:meanvar}
W^{(F)}_O(\mb{s}_i,t_j)=\mu^{(H)}_O(\mb{s}_i,t_j)+\frac{\sigma_{O}^{(H)}(\mb{s}_i)}{\sigma_{S}^{(H)}(\mb{s}_i)}\left\{W^{(F)}_S(\mb{s}_i,t_j)-\mu^{(H)}_S(\mb{s}_i,t_j)\right\},
\end{equation}
where $\sigma_{\ell}^{(H)}(\mb{s}_i)$ for $\ell=\{O,S\}$ is the standard deviation in the historical period, and is estimated from the parameters of the autoregressive process in \eqref{eq:trend}.

\subsection{Covariance adjustment}\label{sec:bicov}

In the previous section, the adjustment was performed independently for every grid point, hence focusing only on the marginal distribution, without considering the potential dependence across multiple locations. However, there is strong evidence of spatial dependence in the observed and simulated field. Indeed, Figure S10 indicates a strong empirical spatial correlation between two neighboring points for both data sets. It is therefore of interest to develop methods to correct not just for the mean and variance, but also for the spatial correlation. 

A simple generalization of \eqref{bias:meanvar} allows for adjustment of both the mean and covariance matrix (therefore including variance), and under the assumption of Gaussianity for both observations and simulations this would be sufficient to fully characterize a transformation for the joint distribution. Indeed, future observations can be expressed as:
\begin{equation} \label{bias:meancov}
\mb{W}^{(F)}_O (t_j)=\bs{\mu}^{(H)}_O(t_j)+(\bs{\Sigma}_{O}^{(H)})^{1/2}\left\{(\bs{\Sigma}_{S}^{(H)})^{-1/2}\right\}^\top\left\{\mb{W}^{(F)}_S (t_j)-\bs{\mu}^{(H)}_S(t_j)\right\},
\end{equation}
where $\bs{\mu}^{(H)}_{\ell}(t_j)=\left(\mu^{(H)}_{\ell} (\mb{s}_1,t_j),\ldots, \mu^{(H)}_{\ell} (\mb{s}_n,t_j)\right)^\top$ (with the same convention being used for the other quantities), and $\bs{\Sigma}_{\ell}^{(H)}$ for $\ell=\{O,S\}$ denotes the covariance matrices for the observational and simulated data sets in the historical period, and the superscript $1/2$ denotes the Cholesky decomposition. While $\bs{\Sigma}_{\ell}^{(H)}$ could be estimated using a simple nonparametric sample covariance matrix, the relatively large number of locations ($n=614$) would result in an unstable estimate. Instead, we assume spatial dependence via two parametric models in increasing order of complexity. First, we consider a stationary and isotropic covariance function modeled by the Mat\'ern correlation \citep{ste99}, i.e., if two measurement are separated by distance $h$, then their correlation is 
\begin{equation}\label{eq:mat}
C(h)=\frac{2^{1-\nu}}{\Gamma(\nu)}\left(\sqrt{2\nu}\frac{h}{\rho}\right)^{\nu}K_{\nu}\left( \sqrt{2\nu}\frac{h}{\rho}\right),
\end{equation}
where $\rho>0$ is the range parameter, $\nu>0$ is the smoothness of the process, and $K_\nu$ is the Bessel function of the second kind of order $\nu$. Second, we consider a more articulated nonstationary covariance resulting from the kernel convolution representation of the random field \citep{hig02}, which can be written in closed form under a general class of kernels as \citep{pac06}
\begin{equation}\label{eq:ns}
C(\mb{s},\mb{s}')=\sigma(\mb{s})\sigma(\mb{s}')\frac{|\bs{\Sigma}(\mb{s})|^{1/4}|\bs{\Sigma}(\mb{s'})|^{1/4}}{\left|\frac{\bs{\Sigma}(\mb{s})+\bs{\Sigma}(\mb{s}')}{2}\right|^{1/2}}g\left(\sqrt{Q(\mb{s},\mb{s}')}\right),
\end{equation}
where 
\[
Q(\mb{s},\mb{s}')=(\mb{s}-\mb{s}')^\top \left(\frac{\bs{\Sigma}(\mb{s})+\bs{\Sigma}(\mb{s}')}{2}\right)^{-1}(\mb{s}-\mb{s}'). 
\]
In this study, the function $g$ is specified to be the exponential function (although several other alternatives are possible) with a spatially varying range. All the spatially varying parameters are defined through the mixture of fixed knots for $A$ fixed locations $\mb{b}_a, a= 1,\ldots,A$. Herein, we select $A=4$ as a larger number of knots would imply a considerable increase in computational cost. The spatially varying variance is defined as
\[
\sigma(\mb{s})=\sum_{a=1}^A w_a(\mb{s})\sigma_a, \quad  w_a(\mb{s})\propto \exp \left\{-\frac{\|\mb{s}-\mb{b}_a\|^2}{2\lambda_{\sigma}}\right\},
\]
where the weights are normalized and $\lambda_{\sigma}$ could be estimated, but for simplicity it is here assumed to be fixed at one half of the minimum distance between the knots. A similar approach is used to define $\bs{\Sigma}(\mb{s})$ and the exponential function range. Inference is performed through local likelihood, see \cite{ris17} for details. 

\section{Adjusting for non-Gaussian spatial data}\label{sec:mod}

\begin{figure}[h]
\centering
\includegraphics[scale=0.5]{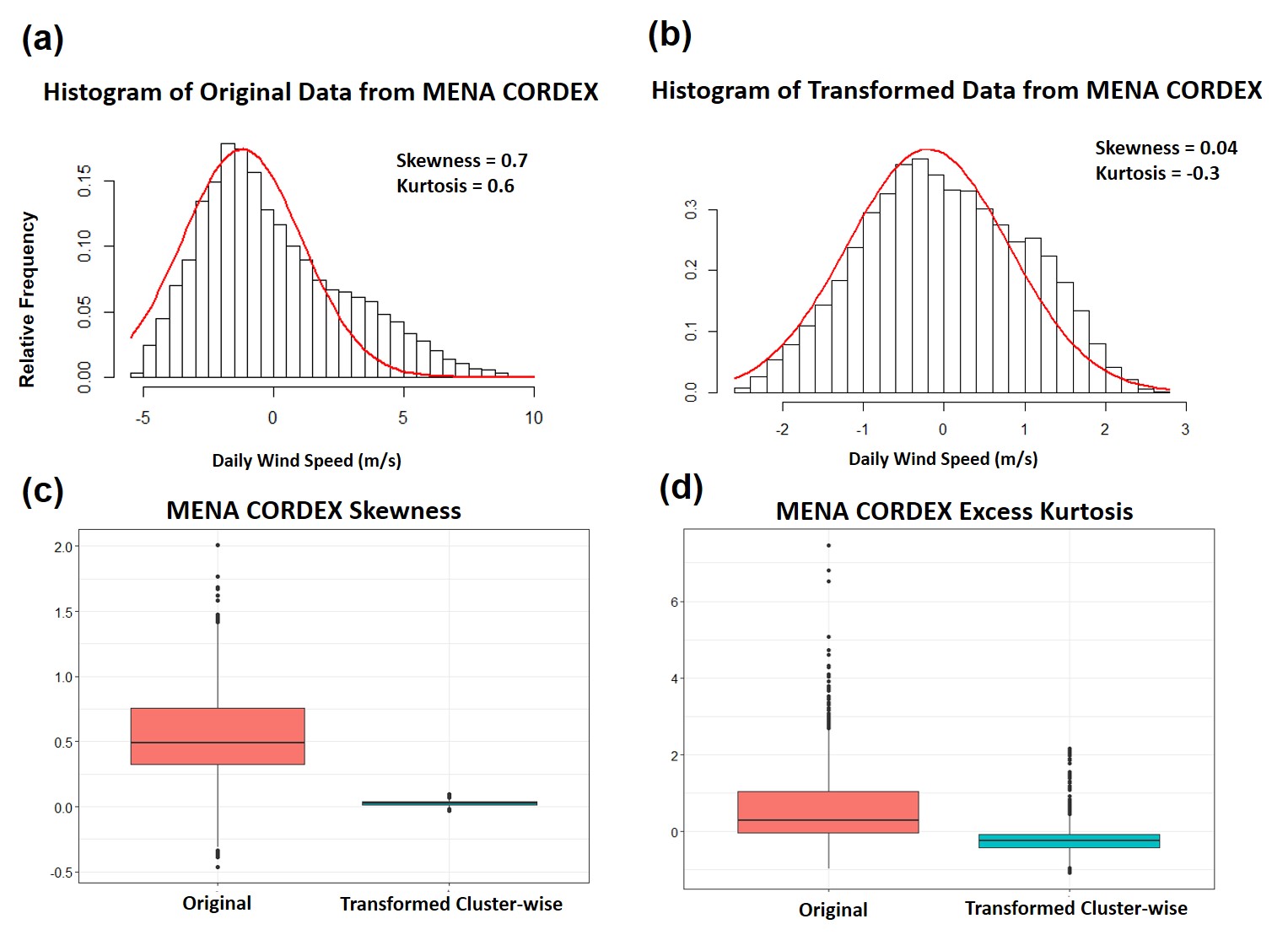}
\caption{Comparison of the MENA CORDEX data from 1980 to 2005 before and after transformation. Histogram of the (a) original and (b) transformed data at one selected location (see the cross in Figure S1(a-b)). The red line represents the Normal distribution that best fits the data. Boxplot of (c) skewness and (d) excess kurtosis at all locations.}
\label{fig:hist}
\end{figure}

The methods presented in Section \ref{sec:rev} allow to correct the joint distribution under the assumption that the mean and covariance are the only quantities for which the adjustment is necessary. This is appropriate only for physical variables whose temporal aggregation is sufficiently large to ensure at least approximate Gaussianity. Wind data aggregated at daily or subdaily level are generally expected to exhibit non-Gaussian behaviors, and in particular, to be skewed to the right owing to occasionally high values because of local meteorological events such as storm fronts or persistent wind gusts. This is indeed the case in our application. Figure \ref{fig:hist}(a) shows the histogram of the MENA CORDEX data from a location at the northwest end of Saudi Arabia represented by a cross in Figure S1(a-b). The superimposed best fit for a Gaussian distribution is vastly inadequate as it fails to capture the aforementioned right skew. Furthermore, the red boxplots in Figure \ref{fig:hist}(c-d) show the skewness and excess kurtosis for all locations in MENA CORDEX, with a characteristic right skew and some degree of excess kurtosis.

We propose a correction method for non-Gaussian data based on marginal, spatially varying transformations. We denote this transformation to Gaussianity by $g_{\lambda}$, a function depending on $\lambda$, and we assume the same class of transformation by $g_{\bs{\lambda}}$ with  $\bs{\lambda}=(\lambda_1,\ldots,\lambda_n)^\top$ indicating the element-wise transformation at each location. Future observations are then obtained as
\begin{equation} \label{bias:nongau}
\mb{W}^{(F)}_O(t_j)=g^{-1}_{\bs{\lambda}_O}\left[\bs{\mu}^{(H)}_O(t_j)+(\bs{\Sigma}_{O}^{(H)})^{1/2}\left\{(\bs{\Sigma}_{S}^{(H)})^{-1/2}\right\}^\top\left\{g_{\bs{\lambda}_S}(\mb{W}^{(F)}_S)(t_j)-\bs{\mu}^{(H)}_S(t_j)\right\}\right],
\end{equation}
where now $\bs{\mu}^{(H)}_{\ell}(\mb{s}_i)$ and $\bs{\Sigma}_{\ell}^{(H)}$ for $\ell=\{O,S\}$ represent the mean and covariance matrix of the \emph{transformed} process $g_{\bs{\lambda}_{\ell}}(\mb{W}^{(H)}_{\ell})(t_j)$, respectively. In the next sections, we will discuss the particular choice of transformation $g_{\bs{\lambda}}$ and the assumptions on $\bs{\lambda}$ to allow an approximately optimal choice across the entire spatial domain. 

\subsection{Relative merits of the proposed approach and the Kennedy and O'Hagan framework}

The methodology we propose in this work bears some similarities but also differences with a widely used framework for coupling data and simulations initially proposed by \cite{ken00} and \cite{ken01}. Indeed, their model can be written as
\begin{equation}\label{eq:cal}
W_O(\mb{s})=\rho \cdot W_S(\mb{s})+\delta(\mb{s})+\epsilon(\mb{s}),
\end{equation}
where the subscripts \textit{O} and \textit{S} indicate observations and simulations, respectively. In its simplest form, this is a deterministic correction with $\rho$ and $\delta(\mb{s})$ which is identifiable as long as $\rho$ is constant \citep{tuo15}. Equation \eqref{bias:meancov} resembles this expression, but it is more general as it proposes a linear transformation of the entire spatial field. Indeed, the closest equivalent in the \cite{ken00} framework would be 
\begin{equation}\label{eq:calv}
\mb{W}_O=\bs{\rho} \cdot \mb{W}_S+\bs{\delta}+\bs{\epsilon},
\end{equation}
where $\mb{W}_O, \mb{W}_S, \bs{\delta}$ and $\bs{\epsilon}$ are $n$-dimensional vectors, and $\bs{\rho}= \rho {\bf I}_n$ is an $n\times n$ matrix. Our approach as detailed in Equation \eqref{bias:nongau} is more general, as it does not just assume a linear transformation, but also a nonlinear function for both observations and simulations to achieve Gaussianity and hence theoretically justifies our transformation. It is therefore more similar in spirit to some of the most recent work in calibration using nonlinear functions such as deep neural networks \citep{bha20}. 

Another important difference is that while \eqref{eq:cal} articulates a stochastic relationship, our model is instead deterministic, as the adjustment parameters are chosen by simple linear algebra operations to transform a Normal vector into another one, and a (nonlinear) transformation to normality. Our choice of a deterministic relationship was mostly justified by the need to align our work with the current practice in bias correction.  

On the computational side, the choice of our deterministic transformation has the advantage of requiring an estimation of parameters from observations and simulations separately. In the context of our problem, where the observational data set has $n\times T=614 \times 9,497\approx 5.8\times 10^6$ observations, this is a very desirable feature especially for a nonstationary spatio-temporal model already considerably time consuming such as the one we propose.

Finally, our work focuses on bias correcting observations with a computer simulation such as MENA CORDEX for which we do not have control over the input parameters. This approach is extremely common in geoscience, and a wide range of literature can be found, mostly focused on Earth System Models. The approach by \cite{ken00,ken01} is generally applied for calibration of computer models, i.e., determining the best choice of numerical model input so that the simulations would resemble the observations. Calibration is typically performed assuming tens, or even hundreds of simulations are available for a given problem, with input specified from some design criterion such as a Latin hypercube design. This number is simply not achievable here, we only have five MENA CORDEX simulations with an extremely large input space, so in \cite{che18} we chose the best one according to some in-situ observations.

\subsection{The Yeo-Johnson transformation}\label{yj-trans}

The standard Box-Cox transformation \citep{erdin} cannot be used for $g_{\lambda}$, as our objective is to transform residuals, which are not necessarily non-negative. Instead, we rely on a similar function, the Yeo-Johnson transformation \citep{yeo00}:
\[
g_{\lambda}(x)=\left\{
\begin{array}{ll}
      [(x+1)^\lambda-1]/\lambda, & x \geq 0, \lambda \neq 0, \\
      \log(x+1), & x \geq 0,\lambda=0, \\
      -[(-x+1)^{2-\lambda}-1]/(2-\lambda), & x < 0, \lambda \neq 2,\\
      -\log(-x+1), & x<0,\lambda=2.\\
\end{array} 
\right. 
\]

While the standard approach for estimating $\lambda$ is to maximize the likelihood of transforming the data to Gaussianity, in this work we propose a different inference approach for both observation and simulation so that their (joint) distributional distance, which in this work is measured as the Kullback-Leibler divergence (KL divergence \citep{kul51}, see Section \ref{sec:kldiv}), is minimized. The key idea of our proposed approach is that even if the parameter is estimated by minimization of KL divergence and not Maximum Likelihood Estimator (MLE), their difference is not substantial. Indeed, a sizable number of sites ($50\%$ and $37\%$ for observations and simulations, respectively), have cluster-wise transformation parameters within the (asymptotic, likelihood based) 95\% MLE confidence interval including the parameter obtained via distance minimization, despite its narrow range owing to the sample size from $T=9,497$ days (see Figure S11). Once the parameter has been inferred for both data sets and the transformation has been applied, the bias and covariance from the correction method in Section \ref{sec:bicov} are estimated under the assumption of Gaussianity, and the results are applied to the future simulated data to obtain an estimate for future observational data as explained in equation \eqref{bias:nongau}. As an example, Figure \ref{fig:hist}(b) shows the result for the transformed wind field according to the method proposed in this study for one location, resulting in an approximately Gaussian distribution. Figure \ref{fig:hist}(c-d) show the skewness and excess kurtosis at all the locations before and after the transformation (with MLE, see Figure S12 for the results with KL divergence), and it is readily apparent how the transformation is sufficiently flexible to transform the data to Gaussianity at all locations.  

\subsection{$k$-Nearest Neighbors Approximation of the Kullback-Leibler Divergence}\label{sec:kldiv}
 
The KL divergence is a measure of how one probability distribution differs from a second reference probability distribution. In this study, we estimate the KL divergence between the simulations and observations, denoted as $f_S$ and $f_O$ respectively:
\begin{equation} \label{eq:kl}
D_{KL}(f_O||f_S)=\int f_O(x)\log\left\{\frac{f_O(x)}{f_S(x)}\right\}\mathrm{d}x.
\end{equation}
In our case, a direct evaluation is computationally impossible as comparing two multivariate distributions of dimension $n=614$ would require an integration over the same number of dimension in \eqref{eq:kl}. While  a simplifying expression is available for the Gaussian distribution, it will not be used here as a comparison with different methods in the original non-Gaussian scale must be made. Therefore, we rely upon a numerical approximation of the integral using $k$-nearest-neighbor ($k$-NN) \citep{li,wang}. Let us assume that we have two $n$-dimensional random samples from two populations with probability density functions $f_O$ and $f_S$. We define the distance between the $i$th sample from $f_O$ and 1) its $k$-NN in the same population as $\rho_{k'}(i)$, and 2) its $k$-NN in the other population as $\nu_{k'}(i)$. Then, we use the following approximation \citep{wang}:
\begin{equation} \label{eq:lln}
\hat{D}_{KL}(f_O||f_S)=\frac{1}{n}\sum_{i=1}^{n}\log\frac{\hat{f}_O(\bs{X}_i)}{\hat{f}_S(\bs{X}_i)}=\frac{n}{m}\sum_{i=1}^{m}\log\frac{\nu_{k'}(i)}{\rho_{k'}(i)}+\log\frac{m'}{m-1},
\end{equation}
where $\hat{f}_{O}^{(k')}(\bs{X}_i)=\frac{k'}{m-1}\cdot\frac{1}{v_1(n)\rho_{k'}^d(i)}$ and $\hat{f}_{S}^{(k')}(\bs{X_i})=\frac{k'}{m'}\cdot\frac{1}{v_1(d)\nu_{k'}^d(i)}$ are the $k$-NN estimators of $f_O$ and $f_S$, respectively, $v_1(n)=\frac{\pi^{n/2}}{\Gamma(\frac{n}{2}+1)}$,
and $\Gamma(\cdot)$ is the Gamma function. 

In our case, the dimension is equal to the number of locations, i.e. $n=614$, and the number of samples $m=m'=4,749$ (13 years of data in the testing set, as detailed in Section \ref{sec:prepr}) since we evaluated MENA CORDEX and MERRA-2 for the same time period. One critical aspect of this approximation is the selection of $k'$, i.e., the number of nearest neighbors. In this study, the standard convention of using the square root of the number of observations is used \citep{boltz}, so that $k'=\sqrt{4,749}\approx 70$. 

\subsection{K-means Clustering}\label{sec:kmeans}

To determine the transformation in \eqref{bias:nongau}, $\bs{\lambda}_O$ and $\bs{\lambda}_S$ must be estimated. While it is possible in principle to estimate all $2n=1,222$ location-specific parameters jointly, this would lead to an unnecessary and severe overparametrization, as geographically close locations are expected to have similar estimates. In practice, a joint estimation is also computationally infeasible, since it would require a simultaneous optimization over all $2n$ parameters. 

In order to reduce the parameter space, and consequently the computational time, we assume that $\bs{\lambda}_O$ and $\bs{\lambda}_S$ are constant across some regions with a $k$-means clustering approach, i.e., for some $p$-variate observations $\mb{x}_1,\ldots, \mb{x}_n$ and a fixed $k''$, we aim to minimize the distance between elements of each cluster and its mean.

\begin{figure}[ht!]
\includegraphics[scale=0.5]{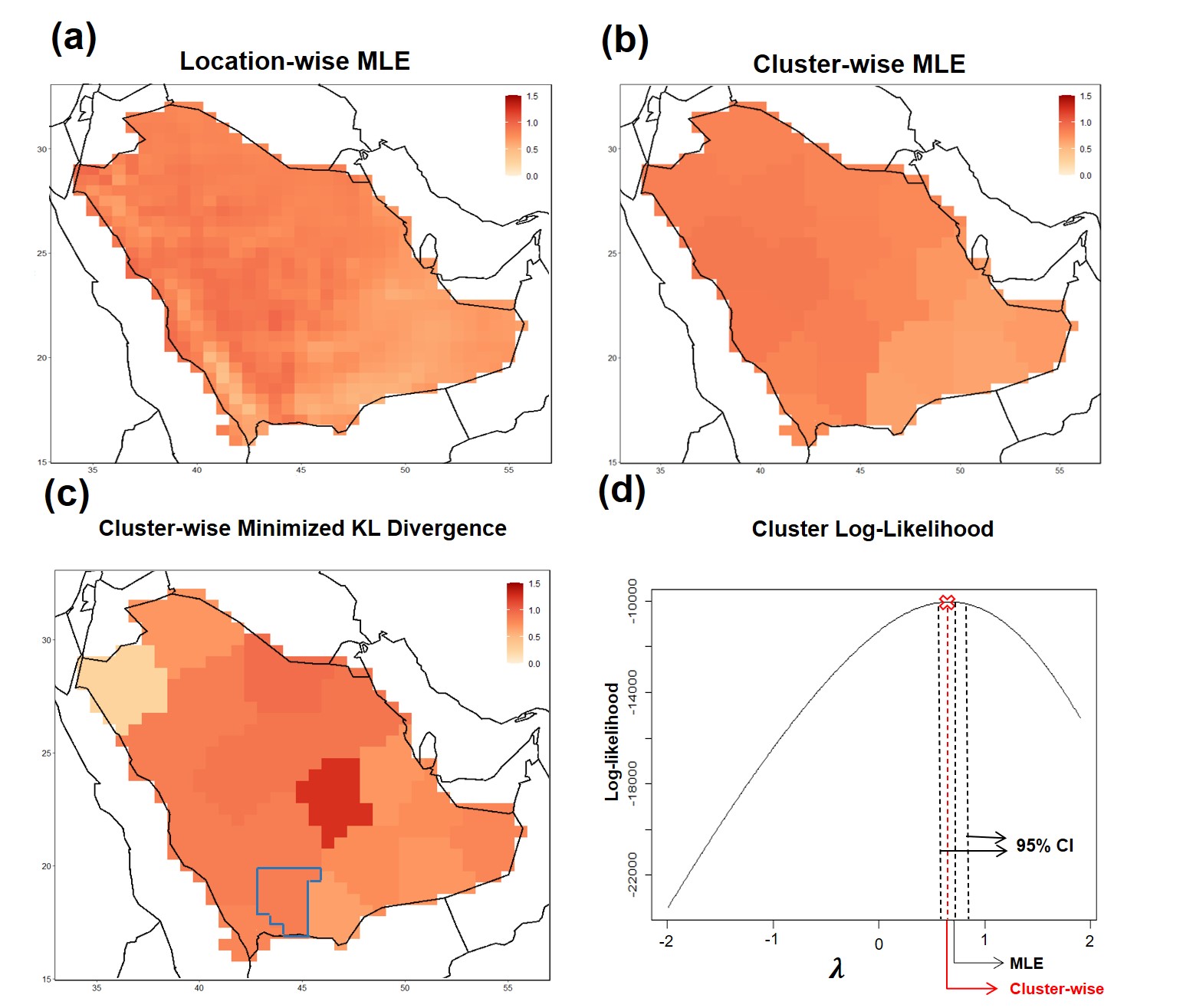}
\caption{MENA CORDEX maps of (a) the Yeo-Johnson MLE; (b) the cluster-wise MLE and (c) the cluster-wise parameter estimated by minimizing the KL divergence. (d) The log-likelihood profile of the cluster indicated in (c), the red cross represents the cluster-wise parameter, and the outer-most dashed lines indicate the 95\% confidence interval of the MLE.}
\label{fig:trans}
\end{figure} 

While our main interest is to obtain regions with similar $\bs{\lambda}_O$, it would be desirable to have spatially coherent clusters for interpretability. Therefore, we assume that for each site, we have $p=3$ covariates: the MLE of the element of $\bs{\lambda}_O$ corresponding to that site, the latitude, and the longitude. Preliminary exploratory analysis (not shown) indicates that a weighted version of $k$-means with even a small weight on latitude and longitude allows for spatially coherent clusters. Therefore, we will assume a weight of 0.98 on the MLE and a weight of 0.01 each for the latitude and longitude, and $k''=20$ clusters, each with thirty to fifty grids, as shown in Figure \ref{fig:trans}(b-c). The usage of clusters instead of location-specific parameters implies some loss of information. For MENA CORDEX, Figure \ref{fig:trans}(a) shows the map of the pointwise MLEs (see Figure S10(a) for MERRA-2), whereas Figure \ref{fig:trans}(b) shows the clusterwise MLEs. Both the values and spatial patterns are visually similar. Figure S10(b-c) show the map of the points whose (asymptotic, likelihood-based) confidence interval for pointwise MLEs include the MLE for the corresponding cluster of both data sets. Despite the very small intervals resulting from $T = 9,497$ observations, a sizable portion of the sites (50\% MERRA-2 and 37\% for MENA CORDEX) includes the cluster MLE. 

Even though our objective is to estimate  $\bs{\lambda}_{\ell}=(\lambda_{\ell;1},\ldots,\lambda_{\ell;k''})^\top, \ell=\{O,S\}$ (i.e., the parameters for each cluster) such that the KL divergence is minimized, the resulting value must not be too different from the MLE to still achieve Gaussianity. Figures \ref{fig:trans}(b) and \ref{fig:trans}(c) compare the cluster-wise MLE and KL minimizer for MENA CORDEX. Their patterns are very similar, albeit with a few noticeable differences. 12 clusters from among $k''=20$ have the KL minimizer inside the MLE confidence interval for both MENA CORDEX and MERRA-2, and Figure \ref{fig:trans}(d) shows an example of the cluster highlighted in the southern part of the country. In the case of pointwise MLEs, despite the narrow interval from the large number of observed days, the KL minimizers in panel (c) fall within the interval from panel (b). 

\section{Validation with simulations and real data}\label{sec:valid}

In this section, we validate our proposed methodology with respect to traditional approaches in terms of the KL divergence for 1) two simulated non-Gaussian random fields, and 2) the MENA CORDEX simulations and MERRA-2 data. Throughout this section, we use M to denote the marginal bias correction with mean \eqref{bias:mean}, MV to denote mean and variance \eqref{bias:meanvar}, MC and MN to denote the mean and covariance in model \eqref{bias:meancov} via the Mat\'ern parameters in \eqref{eq:mat} and the nonstationary model \eqref{eq:ns}, respectively. Further, we compare these methods with our proposed approach in \eqref{bias:nongau} for two cases in which the nonstationary covariance \eqref{eq:ns} is used. In the first case, denoted as T1, a single transformation parameter is considered for each of the observations and simulations, so that all locations are transformed by the same parameter, minimizing the KL divergence. In the second case denoted as TC, we use a cluster-wise parameter determined by $k$-means as specified in Section \ref{sec:kmeans}, for each of the two data sets. Thus, the locations in the same cluster are transformed by the same parameter and the vector of all cluster transformations minimizes the KL divergence across all the sites. Computational details are presented in the supplementary material. 

\subsection{Simulated data}

We perform a simulation study to assess the model performance in capturing varying degrees of non-Gaussianity. Both data sets are simulated from two random fields that cannot be transformed to Gaussianity with a simple marginal transformation. Furthermore, to simplify the setting, we do not assume any inter-annual or annual trend or temporal dependence (i.e., $\phi_{i',i}=\beta_{k,i}=\beta'_{k,i}=\omega_i=0$ for all $i=1,\ldots, n, i'=1,\ldots, P$, and $k=1,\ldots, K$ in \eqref{eq:trend}). The spatial domain of our simulation comprises of 200 locations divided into $R=8$ spatial regions with 25 points each on a $5\times 5$ grid in a square of length 1. For each simulation, 100 replicates were generated, among which the first 50 were considered the historical period, and the last 50 were considered the future period. A total of 1,000 simulations was performed. 

As `observational' data, we simulated samples from a bi-resolution model with a non-Gaussian marginal distribution \citep{tag21,tag20b}. For each region $r=1,\ldots, R$ we have 
\begin{equation}\label{eq:skewt}
W^{(\ell)}_{O}(\mb{s})=\frac{\lambda^{\text{SKT}} |U^{\text{SKT}}_r|+\eta^{\text{SKT}}_r(\mb{s})}{\sqrt{Z^{\text{SKT}}_r}},
\end{equation}
\begin{sloppypar}
\noindent where $\ell=\{H,F\}$ for historical and future time period, and the temporal index has been omitted for simplicity, $Z^{\text{SKT}}_r\sim \text{Gamma}(\nu^{\text{SKT}}/2,\nu^{\text{SKT}}/2)$ independent and identically distributed, and $\mb{U}^{\text{SKT}}=(U^{\text{SKT}}_1,\ldots,U^{\text{SKT}}_R)^\top\sim \mathcal{N}(\mb{0},\bs{\Sigma}^{\text{SKT}}_0)$ and $\eta^{\text{SKT}}_r(\mb{s})$ indicate a zero mean Gaussian random field independent across $r$ with covariance matrix $\bs{\Sigma}^{\text{SKT}}_r$. The model assumes that, for each region $r$, the two effects $Z^{\text{SKT}}_r$ and $U_r^{\text{SKT}}$ are constant, and the small scale variation is accounted by the field $\eta^{\text{SKT}}_r(\mb{s})$. Across the regions, the vector $\mb{U}^{\text{SKT}}$ characterizes the large-scale dependence. This model \eqref{eq:skewt} exhibits a skew-$t$ marginal distribution, i.e., a perturbation of the $t$ distribution accounting for skewed behavior \citep{azz03} and it can be represented hierarchically, thereby allowing relatively fast frequentist or Bayesian inference. For our simulation, we fix the parameters $\lambda^{\text{SKT}}=0.8$ and $\nu^{\text{SKT}}=8$ and assume that $\bs{\Sigma}^{\text{SKT}}_r$ and $\bs{\Sigma}^{\text{SKT}}_0$ are generated from  exponential covariance functions with ranges of 0.2 and 0.5, respectively, corresponding to maximum correlations of 0.77 and 0.24, respectively. The large correlation within each region is in accordance with the previous results with bi-resolution models, where the majority of the dependence is explained by the small scale \citep{cas18,tag21}. Additional simulations with parameter choices leading to weaker and stronger correlation are shown in Figure S12.
\end{sloppypar}

The `simulation' data are obtained from a Gaussian-Log-Gaussian model (GLG, \cite{pal06}), which assumes 
\begin{equation}\label{eq:glg}
W^{(\ell)}_{S}(\mb{s})=\frac{\eta^{\text{GLG}}(\mb{s})}{\sqrt{\xi^{\text{GLG}}(\mb{s})}}+\varepsilon^{\text{GLG}}(\mb{s}),
\end{equation}
where $\ell=\{H,F\}$ for historical and future time period and $\eta^{\text{GLG}}(\mb{s})$ is an isotropic zero mean Gaussian field with covariance function $C^{\text{GLG}}$, which is considered to be an exponential with a range parameter of 0.2. The process $\xi^{\text{GLG}}(\mb{s})$ is such that $\log\{\xi^{\text{GLG}}(\mb{s})\}$ is a Gaussian field with mean $-\nu^{\text{GLG}}/2$ and covariance function $\nu^{\text{GLG}} C^{\text{GLG}}$, where $C^{\text{GLG}}$ is exponential with a range of 0.7, so that marginally $\xi^{\text{GLG}}(\mb{s})$ is lognormal with mean 1 and variance $\exp(\nu^{\text{GLG}})-1$, and $\nu^{\text{GLG}}=8$ in our simulation. The error term $\varepsilon^{\text{GLG}}(\mb{s})$ represents an independent Gaussian noise with zero mean and variance $\tau^{2;\text{GLG}}=0.1$, representing the micro-scale variability or the nugget effect. As for the bi-resolution model \eqref{eq:skewt}, model \eqref{eq:glg} cannot be transformed to Gaussianity with a simple marginal transformation; thus our proposed approach is not trained to obtain an exact transformation to Gaussianity for either the `observed' or `simulated' data sets.

Figure \ref{fig:result1}(a) shows the results in terms of the KL divergence ratio between the MV method and MC, MN, T1, and TC (M is not applied because the `simulated' data have zero mean by construction), where each element of the boxplots represents one of the 1,000 simulations. Despite the considerably different and non-trivial non-Gaussian structures of the two data sets, the proposed approach can more closely estimate the true future observations against a Gaussian transformation, with an improvement of 29\% in the median using a single transformation in T1. The clusterwise transformation TC yields a further improvement to 31\%, although at an increased computational cost, because the transformation parameters must be separately estimated for each region. 

\subsection{Wind speed data}

Since future observations cannot be used to validate our model, we only consider the historical data period, and we separated it into a training and testing set. We use the first thirteen years of data (1980 to 1992) as the training set and the last thirteen years (1993 to 2005) as the test set. Thus, the models are estimated according to the training set, and the predictions in the test period are made by assuming that only MENA CORDEX is available. Then, MERRA-2 is estimated and compared with the original data. 

We compute the KL divergence (estimated and true MERRA-2 data in the testing set) ratio between the simple mean bias correction (M) and other approaches. The MV, MC, and MN approach result in ratios of 0.93, 0.88, and 0.81, respectively, indicating an increasingly faithful representation of MERRA-2 if the variance and covariance are estimated. Our proposed approach can further decrease the KL ratio to 0.58 and 0.51 for T1 and TC, respectively. 

\begin{figure}
\centering
\includegraphics[scale=0.5]{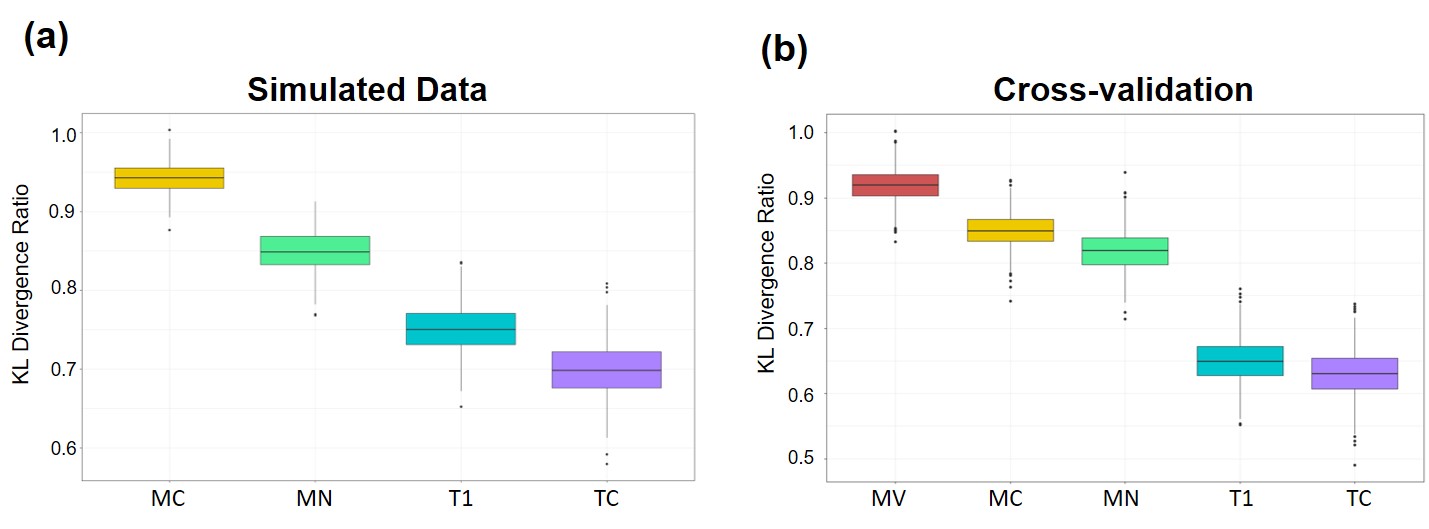}
\caption{Boxplot of KL divergences between (a) MV and the MC, MN, T1 and TC method for the simulated data and (b) M and the MV, MC, MN, T1 and TC method for the estimated and actual MERRA-2 from 1993 to 2005. Each of the 1,000 elements of the boxplot represents an independent simulation in (a) and a random sub-sample of 300 locations in (b). }
\label{fig:result1}
\end{figure}

To assess the uncertainty in the KL ratio, for 1,000 times we sample 300 points (from the total $n=614$) from the spatial domain, perform the correction methods, and estimate the KL ratio for each sample. Each boxplot in Figure \ref{fig:result1}(b) represents the KL ratio between M and the other approaches for each subsample. For MC, MN and T1 methods, we estimate the covariance function parameters for each sample. In the case of the transformation, we used the same parameters as for the full data set for computational convenience. For TC, we used $k$-means and a stratified sampling method to select 50\% of the locations in each cluster. Overall, the improvement in the proposed model is apparent throughout the subsamples, with an increasingly small KL divergence and with our proposed T1 and TC methods yielding the smallest divergence. 

\section{Application}\label{sec:app}

We use the proposed approach to estimate how the daily revenue from the current optimal turbines build-out in Saudi Arabia will be impacted by a changing climate over the next few decades. In subsection \ref{sec:alpha} we describe our approach for extrapolating the predicted surface wind to hub height, whereas in subsection \ref{sec:assch} we assess the final change in profits implied by the corrected data extrapolated at hub height. 

\subsection{Extrapolation to hub height}\label{sec:alpha}

\begin{figure}[ht!]
\centering
\includegraphics[scale=0.6]{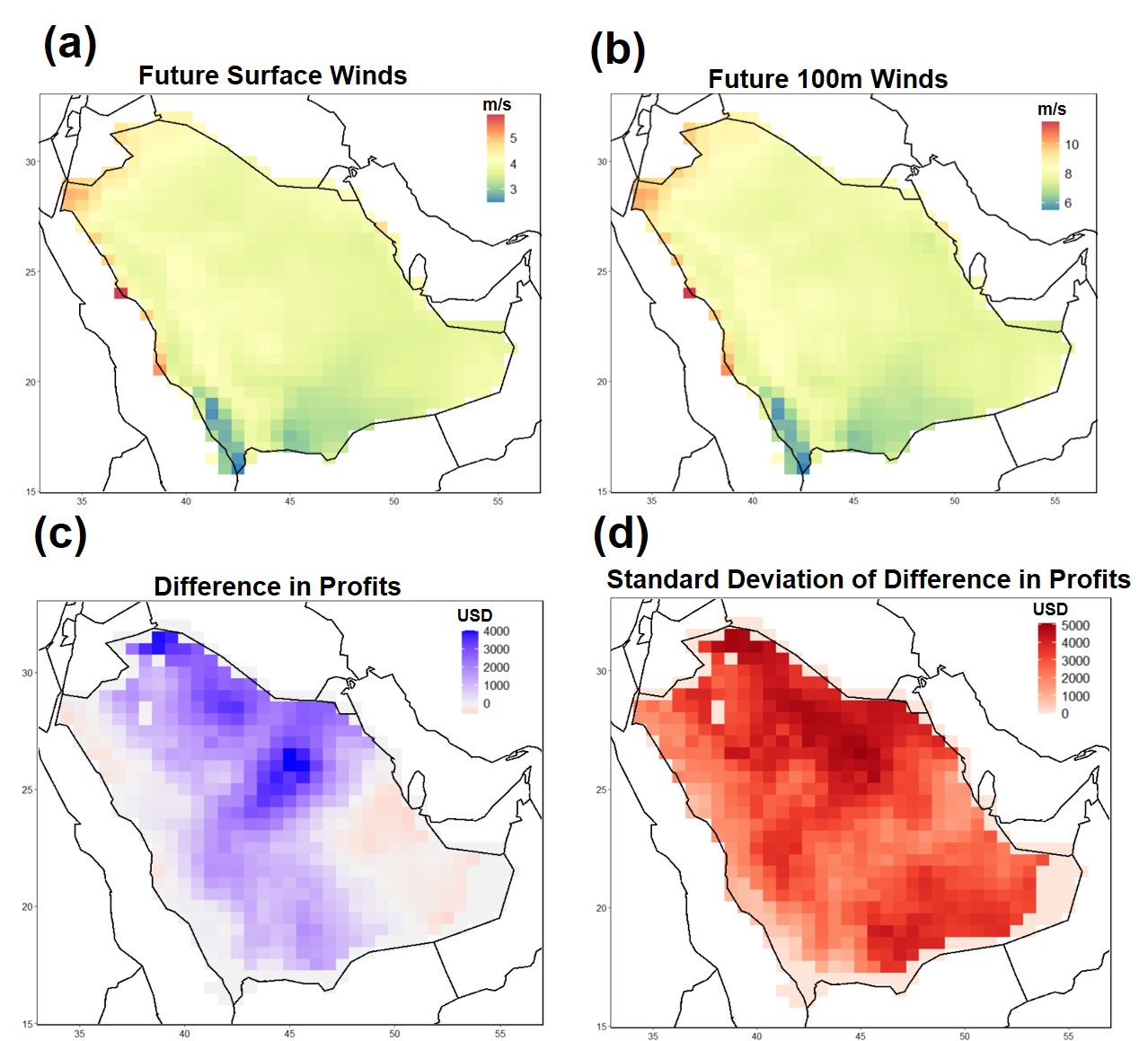}
\caption{Maps of averaged daily wind speed at (a) the surface and (b) 100 meters from 2025 to 2050, and (c) average differences between the future and historical wind energy profits and their standard deviations (d).}
\label{fig:result2}
\end{figure}

In Figure \ref{fig:result2}(a) the daily surface wind speed according to MERRA-2 is presented. This map indicates spatial patterns of potential interest for wind harvesting, especially in the north-west corner, which is a site of particular interest because of the ongoing project to build a self-sustainable city (NEOM, \cite{far19}). However, the surface wind does not necessarily represent the wind at higher altitudes. Indeed, the wind speed data from both MENA CORDEX and MERRA-2 are computed or observed at a reference height of 10 meters, whereas wind turbines normally operate at a height of 80-120 meters. Therefore, in order to assess the wind energy output, the surface wind speed must be extrapolated to the height at which wind turbines operate. There is a vast literature on extrapolating wind speed from surface to a height within the boundary layer, see \cite{emeis} for a comprehensive review. In the vast majority of studies, the power law is used:
\begin{equation} \label{eq:power}
\begin{array}{rcl}
 W_O^{(F)}(\mb{s}_i,t_j,h_k) & = & W_O^{(F)}(\mb{s}_i,t_j,h_r)\left(\frac{h_k}{h_r}\right)^{\alpha_{i,j}}e^{\eta(\mb{s}_i,t_j)},\\[7pt] 
 \eta(\mb{s}_i,t_j) & \sim &  \mathcal{N}(0,\sigma^2_{i}),
 \end{array}
\end{equation}
where $h_k$ is the height to which we want to extrapolate and $h_r$ is the height at which data are available (in our case it is 10 meters). The $\alpha$ is the \textit{shear coefficient} of the power law, and its value is assumed to change depending on local spatial properties such as surface roughness and thermal stability (i.e., the temperature gradient for the first layers in the boundary layer) \citep{gua19}. In the absence of any meteorological information, the standard approach is to assume that $\alpha_{ij}=1/7$, which corresponds to a value observed on flat terrain under neutral atmospheric conditions \citep{rehman, tag19}. Direct estimation of $\alpha_{ij}$ using MERRA-2 or MENA CORDEX is impossible, as direct estimation of the power law from \eqref{eq:power} would require sufficient vertical wind levels below 100 meters, and neither data sets is designed for this level of accuracy near the surface. In this study, we rely instead on a high resolution WRF ensemble and select a run by adopting a planetary boundary layer parameterization, resolution and boundary conditions resulting in simulations closer to a few in-situ data available \citep{gia20}. The WRF simulation is specifically designed to capture the wind at a high resolution near the surface, and resolves the wind speed vertical profile at six levels, which are approximately equally spaced from the ground level to an altitude of 100 meters. Therefore, these data will be used to estimate $\alpha_{ij}$ and $\sigma^2_i$ in \eqref{eq:power}. Since we focus on daily data (neither MERRA-2 nor MENA CORDEX has hourly data) and a preliminary analysis (not shown) did not highlight temporal changes in the shear coefficient, $\alpha_{ij}=\alpha_{i}$ is estimated from the power law with a simple log regression. 

The map of the estimated wind shear coefficients, along with their standard deviations and the coefficients of determination $R^2$ can be observed from Figure S13. The map of the estimates indicates a considerable spatial variability, with very small and even negative values of the shear coefficient closely associated with the mountainous areas of the Hejaz region in the west. While slower wind at high altitudes is an unexpected (yet physically admissible) behavior within the boundary layer, it often occurs in areas exhibiting the smallest $R^2$ and the largest standard deviation, indicating that the power law \eqref{eq:power} is likely not an appropriate model, as also indicated in many studies \citep{gua19,cri20}. Since the installation of wind turbines is not cost effective in the rough terrains of Saudi Arabia \citep{gia20}, the model misspecification over these regions is not a major concern. Once the $\alpha_i$ are estimated, MERRA-2 data are then downscaled from the 50$\times$50  km to the 6$\times$6 km WRF resolution with ordinary kriging using another Mat\'ern \eqref{eq:mat} covariance, and the wind speed at the desired hub height is computed. Finally, given the uncertainty associated with the determination of the shear coefficients, we performed 1,000 extrapolations for each site by simulating both the shear coefficient with its variability, as well as the random noise from the parameters estimated in \eqref{eq:power}.

Figure \ref{fig:result2}(b) shows the extrapolated average daily wind at 100 meters for MERRA-2 during 2025-2050. Besides being uniformly faster than wind at the surface, the spatial patterns are not drastically different from those in panel (a), as expected. Figure S14 shows the relative change of 100 meter wind against surface wind, and highlights how the increase in wind speed ranges from 60\% to about 85\%.

\subsection{Assessing changes in future wind energy revenues}\label{sec:assch}

\cite{gia20} showed that the most cost effective choice of turbines depends upon the site, because the construction and maintenance cost as well as the potential for harnessing the wind vary. Here, we use the same map of the most efficient turbines as indicated in Figure 5(a) of the aforementioned manuscript, and extrapolate the surface wind to the hub height of each turbine type. The wind speed is translated into energy through a turbine-specific power curve, which is a function that represents the ability of the turbine to translate the blade movement into energy. Power curves are zero until a given wind speed, increase up to a maximum rated power and are constant for any stronger wind. We make use of of proprietary database purchased from \textit{The Wind Power} (www.thewindpower.net/about$\_$en.php) providing the nominal power curves for the most popular turbines worldwide. The actual power curves are generally not available and require \textit{ad hoc} methodology for their estimation \citep{yu19}. The total energy output per cell can be calculated by multiplying the power for a single turbine with the total number of turbines, is dependent on the diameter of the rotor blade and has to allow for sufficient spacing among turbines to avoid local turbulence (wake effect), which would reduce the overall efficiency. 

\begin{sloppypar}
Finally, the wind power is translated into actual revenue using the current tariffs of the Saudi Electricity Company (www.se.com.sa/en-us/customers/Pages/TariffRates.aspx), which is approximately 5 US cents per kWh. The current number is likely to be a sensible overestimation because the associated direct and indirect costs were not accounted for. The difference in revenue, as implied by the changes between future and present wind, is presented in Figure \ref{fig:result2}(c). Overall, the map indicates increased revenues of up to approximately \$4,000 per day for the vast majority of the sites. Most importantly, the sites with increased future revenues are those located near the coasts, hence with more potential for wind harvesting, and especially the NEOM region in the north west. The regions towards the center of Saudi Arabia, including the neighborhoods of the capital Riyadh would incur a loss in revenue from changes in wind, if turbines were to be installed there, although it would not be as substantial as in the other areas. The uncertainty map is presented in Figure \ref{fig:result2}(d), and indicates large uncertainties in some areas, especially near the Persian Gulf. While these uncertainties are arguably large in some areas, they are a direct consequence of the considerable daily variability of the power law \eqref{eq:power}. The uncertainty in the use of the power law has been acknowledged \citep{gua19,cri20}, even though it has not been frequently used with the goal of uncertainty quantification, but in relation to additional covariates such as temperature gradients and stability metrics, which cannot be determined in this study given the coarse vertical structures of MERRA-2 and MENA CORDEX. 
\end{sloppypar}

While the aforementioned maps provide an overall estimation of the difference in profits across the country, the areas of highest interest are the ones where the installation of wind turbines would be the most cost-effective choice. Therefore, we estimate the difference in profits for the 75 wind farms locations identified in \cite{gia20} to be the most promising sites. We add the difference in profits across the sites and we obtain a total increase in profit of approximately \$272,000 and a standard deviation of \$14,000, hence lending additional support for the long-term profitability of the selected sites.  In Figure S15 we have recalculated the daily profits using a simple MV approach, i.e., by adjusting only mean and variance in the optimal sites, and we obtained a considerably decreased daily profit of \$111,000 (standard deviation \$26,000).

\section{Conclusion}\label{sec:conc}

In this study, we have addressed the issue of resilience of Saudi Arabia's current plan to diversify its energy portfolio with wind energy under changing climate conditions. The key element associated with this assessment is the estimation of future winds using reanalysis data. In order to provide that, we proposed a novel trans-Gaussian cluster-wise adjustment model based on minimizing the KL divergence between reanalysis and simulated data. 
Once the model was properly validated with a simulation study and with historical data, we estimated future winds from 2025 to 2050. These estimates were extrapolated to the turbine hub height using the output from high resolution numerical model simulations specifically designed to capture the vertical structure in the boundary layer. Finally, these estimates were translated into wind power and changes in revenue from future to present winds. Results show a sizable increase of approximately \$273,000 in daily revenues at the ideal construction sites, albeit with a standard deviation of \$15,000 due to the uncertainty propagated by the extrapolation of wind from the surface to hub height. 

From a methodological perspective, our model generalized some of the most common approaches used for correcting spatial fields to the non-Gaussian case and proposed an inferential approach that could be scaled to considerably higher spatial resolutions for future data sets with 1) clustering of the trans-Gaussian transformation and 2) numerical approximation of the KL divergence with $k$ nearest neighbors. The limitations of the current approach will likely become apparent if hourly or sub-hourly resolution were made available: in that case, the space-time interaction could not be reduced to a simple vector moving average model and would likely require nonseparable models, the application of which would considerably increase the computational cost. Furthermore, high resolution wind data would 1) result in sparse wind such that a simple trans-Gaussian model could not capture, and a more involved latent Gaussian model would be required and 2) invalidate extrapolation with the power law, which available literature suggests to use only up to hourly resolution. This would prompt the development of more sophisticated nonparametric approaches such as neural networks \citep{vas19}. 

From the applied perspective, even though extrapolation is widely acknowledged to be the most important source of uncertainty associated with the determination of wind energy from surface wind, this study did not account for other sources. First, MERRA-2 is only one of the available reanalysis data products, so further comparison with other products, such as the new ERA5 from the European Centre for Medium-Range Weather Forecasts \citep{her20}, could be performed. Second, power curves have been obtained from proprietary data, but the raw data used to determine them are confidential, and would likely indicate some degree of uncertainty with respect to the determination of the said curve. Third, the translation of power into revenue is highly dependent upon policies and negotiations between the turbine operators and local authorities; therefore this may vary across different regions of Saudi Arabia. Future work will focus on better assessing these sources of uncertainty by engaging both industrial partners and policymakers, and discussing potential changes in the siting work conducted by \cite{gia20} in lights of these new results. 

Finally, we emphasize how the scope of this study could be generalized to any country with an emerging wind energy portfolio. Even though vertical extrapolation required a high resolution ensemble focused on Saudi Arabia and energy costs are expected to change across different countries, the core of our methodology involves only publicly available data and hence can be used to inform strategies for other Gulf countries and beyond. Furthermore, in lieu of new data from a high-resolution ensemble, the power law could be simplified to the case of neutral atmospheric stability and flat terrain, a simplifying assumption widely accepted in wind power literature. 

\bibliographystyle{plainnat}
\bibliography{references}

\begin{thebibliography}{62}
\providecommand{\natexlab}[1]{#1}
\providecommand{\url}[1]{\texttt{#1}}
\expandafter\ifx\csname urlstyle\endcsname\relax
  \providecommand{\doi}[1]{doi: #1}\else
  \providecommand{\doi}{doi: \begingroup \urlstyle{rm}\Url}\fi

\bibitem[Azzalini and Capitanio(2003)]{azz03}
Adelchi Azzalini and Antonella Capitanio.
\newblock Distributions generated by perturbation of symmetry with emphasis on
  a multivariate skew t-distribution.
\newblock \emph{Journal of the Royal Statistical Society: Series B (Statistical
  Methodology)}, 65\penalty0 (2):\penalty0 367--389, 2003.
\newblock \doi{10.1111/1467-9868.00391}.
\newblock URL
  \url{https://rss.onlinelibrary.wiley.com/doi/abs/10.1111/1467-9868.00391}.

\bibitem[Bhatnagar et~al.(2020)Bhatnagar, Chang, Kim, and Wang]{bha20}
S.~Bhatnagar, W.~Chang, S.~Kim, and J.~Wang.
\newblock Computer model calibration with time series data using deep learning
  and quantile regression.
\newblock 2020.
\newblock arXiv:2008.13066.

\bibitem[{Boltz} et~al.(2007){Boltz}, {Debreuve}, and {Barlaud}]{boltz}
S.~{Boltz}, E.~{Debreuve}, and M.~{Barlaud}.
\newblock knn-based high-dimensional kullback-leibler distance for tracking.
\newblock In \emph{Eighth International Workshop on Image Analysis for
  Multimedia Interactive Services (WIAMIS '07)}, pages 16--16, 2007.

\bibitem[{British Petroleum}(2020)]{bp20}
{British Petroleum}.
\newblock {BP} statistical review of world energy, 2020.
\newblock
  www.bp.com/content/dam/bp/en/corporate/pdf/energy-economics/statistical-review/bp-stats-review-2018-full-report.pdf.

\bibitem[Cannon(2018)]{can18}
A.J. Cannon.
\newblock Multivariate quantile mapping bias correction: an n-dimensional
  probability density function transform for climate model simulations of
  multiple variables.
\newblock \emph{Climate Dynamics}, 50:\penalty0 31--49, 2018.

\bibitem[Cannon et~al.(2015)Cannon, Sobie, and Murdock]{can15}
Alex~J. Cannon, Stephen~R. Sobie, and Trevor~Q. Murdock.
\newblock Bias correction of gcm precipitation by quantile mapping: How well do
  methods preserve changes in quantiles and extremes?
\newblock \emph{Journal of Climate}, 28\penalty0 (17):\penalty0 6938 -- 6959,
  2015.
\newblock \doi{10.1175/JCLI-D-14-00754.1}.

\bibitem[Cannon et~al.(2020)Cannon, Piani, and Sippel]{can20}
Alex~J. Cannon, Claudio Piani, and Sebastian Sippel.
\newblock Chapter 5 - bias correction of climate model output for impact
  models.
\newblock In Jana Sillmann, Sebastian Sippel, and Simone Russo, editors,
  \emph{Climate Extremes and Their Implications for Impact and Risk
  Assessment}, pages 77 -- 104. 2020.
\newblock \doi{https://doi.org/10.1016/B978-0-12-814895-2.00005-7}.

\bibitem[Castruccio et~al.(2018)Castruccio, Ombao, and Genton]{cas18}
Stefano Castruccio, Hernando Ombao, and Marc~G. Genton.
\newblock A scalable multi-resolution spatio-temporal model for brain
  activation and connectivity in fmri data.
\newblock \emph{Biometrics}, 74\penalty0 (3):\penalty0 823--833, 2018.
\newblock \doi{10.1111/biom.12844}.

\bibitem[Chen et~al.(2012)Chen, Pryor, and Li]{lian}
Lian Chen, S.~C. Pryor, and Dongliang Li.
\newblock Assessing the performance of intergovernmental panel on climate
  change ar5 climate models in simulating and projecting wind speeds over
  china.
\newblock \emph{Journal of Geophysical Research: Atmospheres}, 117:\penalty0
  D24102, 2012.
\newblock \doi{10.1029/2012JD017533}.
\newblock URL
  \url{https://agupubs.onlinelibrary.wiley.com/doi/abs/10.1029/2012JD017533}.

\bibitem[Chen et~al.(2018)Chen, Castruccio, Genton, and Crippa]{che18}
Wanfang Chen, Stefano Castruccio, Marc~G. Genton, and Paola Crippa.
\newblock Current and future estimates of wind energy potential over saudi
  arabia.
\newblock \emph{Journal of Geophysical Research: Atmospheres}, 123\penalty0
  (12):\penalty0 6443--6459, 2018.
\newblock \doi{10.1029/2017JD028212}.
\newblock URL
  \url{https://agupubs.onlinelibrary.wiley.com/doi/abs/10.1029/2017JD028212}.

\bibitem[Chen et~al.(2021)Chen, Castruccio, and Genton]{che20}
Wanfang Chen, Stefano Castruccio, and Marc~G. Genton.
\newblock Assessing the risk of disruption of wind turbine operations in saudi
  arabia using bayesian spatial extremes.
\newblock \emph{Extremes}, 2021.
\newblock \doi{10.1214/17-AOAS1105}.
\newblock in press.

\bibitem[Crippa et~al.(2021)Crippa, Alifa, Bolster, Genton, and
  Castruccio]{cri20}
P.~Crippa, M.~Alifa, D.~Bolster, M.G. Genton, and S.~Castruccio.
\newblock A heteroskedastic time-varying model for improved hourly wind power
  forecasting.
\newblock 2021.
\newblock under review.

\bibitem[Ding(2019)]{yu19}
Y.~Ding.
\newblock \emph{Data Science for Wind Energy}.
\newblock CRC press, Boca Raton, FL, 2019.

\bibitem[Dunne et~al.(2013)Dunne, John, Shevliakova, Stouffer, Krasting,
  Malyshev, Milly, Sentman, Adcroft, Cooke, Dunne, Griffies, Hallberg,
  Harrison, Levy, Wittenberg, Phillips, and Zadeh]{dun13}
John~P. Dunne, Jasmin~G. John, Elena Shevliakova, Ronald~J. Stouffer, John~P.
  Krasting, Sergey~L. Malyshev, P.~C.~D. Milly, Lori~T. Sentman, Alistair~J.
  Adcroft, William Cooke, Krista~A. Dunne, Stephen~M. Griffies, Robert~W.
  Hallberg, Matthew~J. Harrison, Hiram Levy, Andrew~T. Wittenberg, Peter~J.
  Phillips, and Niki Zadeh.
\newblock {GFDL’s ESM2 global coupled climate–carbon Earth system models.
  Part II: carbon system formulation and baseline simulation characteristics}.
\newblock \emph{Journal of Climate}, 26\penalty0 (7):\penalty0 2247--2267, 04
  2013.
\newblock ISSN 0894-8755.
\newblock \doi{10.1175/JCLI-D-12-00150.1}.

\bibitem[Emeis(2018)]{emeis}
Stefan Emeis.
\newblock \emph{Wind Energy Meteorology - Second Edition}.
\newblock Springer, NY, 04 2018.
\newblock ISBN 978-3-319-72858-2.
\newblock \doi{10.1007/978-3-319-72859-9}.

\bibitem[Erdin et~al.(2012)Erdin, Frei, and Künsch]{erdin}
Rebekka Erdin, Christoph Frei, and Hans~R. Künsch.
\newblock Data transformation and uncertainty in geostatistical combination of
  radar and rain gauges.
\newblock \emph{Journal of Hydrometeorology}, 13\penalty0 (4):\penalty0
  1332--1346, 08 2012.
\newblock ISSN 1525-755X.
\newblock \doi{10.1175/JHM-D-11-096.1}.
\newblock URL \url{https://doi.org/10.1175/JHM-D-11-096.1}.

\bibitem[Farag(2019)]{far19}
A.A. Farag.
\newblock The story of {NEOM} city: Opportunities and challenges.
\newblock In Ibrahim~A. Attia~S., Shafik~Z., editor, \emph{New Cities and
  Community Extensions in Egypt and the Middle East}, pages 35--49. Springer,
  Cham, Switzerland, 2019.

\bibitem[Fran{\c{c}}ois et~al.(2020)Fran{\c{c}}ois, Vrac, Cannon, Robin, and
  Allard]{fra20}
B.~Fran{\c{c}}ois, M.~Vrac, A.~J. Cannon, Y.~Robin, and D.~Allard.
\newblock Multivariate bias corrections of climate simulations: which benefits
  for which losses?
\newblock \emph{Earth System Dynamics}, 11\penalty0 (2):\penalty0 537--562,
  2020.
\newblock \doi{10.5194/esd-11-537-2020}.

\bibitem[Gelaro et~al.(2017)Gelaro, McCarty, Suárez, Todling, Molod, Takacs,
  Randles, Darmenov, Bosilovich, Reichle, Wargan, Coy, Cullather, Draper,
  Akella, Buchard, Conaty, da~Silva, Gu, Kim, Koster, Lucchesi, Merkova,
  Nielsen, Partyka, Pawson, Putman, Rienecker, Schubert, Sienkiewicz, and
  Zhao]{gel:17}
Ronald Gelaro, Will McCarty, Max~J. Suárez, Ricardo Todling, Andrea Molod,
  Lawrence Takacs, Cynthia~A. Randles, Anton Darmenov, Michael~G. Bosilovich,
  Rolf Reichle, Krzysztof Wargan, Lawrence Coy, Richard Cullather, Clara
  Draper, Santha Akella, Virginie Buchard, Austin Conaty, Arlindo~M. da~Silva,
  Wei Gu, Gi-Kong Kim, Randal Koster, Robert Lucchesi, Dagmar Merkova, Jon~Eric
  Nielsen, Gary Partyka, Steven Pawson, William Putman, Michele Rienecker,
  Siegfried~D. Schubert, Meta Sienkiewicz, and Bin Zhao.
\newblock {The Modern-Era Retrospective Analysis for Research and Applications,
  Version 2 (MERRA-2)}.
\newblock \emph{Journal of Climate}, 30\penalty0 (14):\penalty0 5419--5454, 06
  2017.
\newblock ISSN 0894-8755.
\newblock \doi{10.1175/JCLI-D-16-0758.1}.
\newblock URL \url{https://doi.org/10.1175/JCLI-D-16-0758.1}.

\bibitem[Giani et~al.(2020)Giani, Felipe, Genton, Castruccio, and
  Crippa]{gia20}
Paolo Giani, Tagle. Felipe, Marc~G. Genton, Stefano Castruccio, and Paola
  Crippa.
\newblock Closing the gap between wind energy targets and implementation for
  emerging countries.
\newblock \emph{Applied Energy}, 269:\penalty0 115085, 2020.
\newblock ISSN 0306-2619.
\newblock \doi{https://doi.org/10.1016/j.apenergy.2020.115085}.

\bibitem[Gualtieri(2019)]{gua19}
Giovanni Gualtieri.
\newblock A comprehensive review on wind resource extrapolation models applied
  in wind energy.
\newblock \emph{Renewable and Sustainable Energy Reviews}, 102:\penalty0
  215--233, 2019.
\newblock \doi{https://doi.org/10.1016/j.rser.2018.12.015}.

\bibitem[Hawkins et~al.(2013)Hawkins, Osborne, Ho, and Challinor]{haw13}
Ed~Hawkins, Thomas~M. Osborne, Chun~Kit Ho, and Andrew~J. Challinor.
\newblock Calibration and bias correction of climate projections for crop
  modelling: An idealised case study over europe.
\newblock \emph{Agricultural and Forest Meteorology}, 170:\penalty0 19--31,
  2013.
\newblock ISSN 0168-1923.
\newblock \doi{https://doi.org/10.1016/j.agrformet.2012.04.007}.
\newblock Agricultural prediction using climate model ensembles.

\bibitem[Hemer et~al.(2012)Hemer, McInnes, and Ranasinghe]{hemer}
Mark Hemer, Kathleen McInnes, and Roshanka Ranasinghe.
\newblock Climate and variability bias adjustment of climate model-derived
  winds for a southeast australian dynamical wave model.
\newblock \emph{Ocean Dynamics}, 62\penalty0 (1):\penalty0 87--104, 2012.
\newblock ISSN 16167341.
\newblock URL
  \url{http://search.ebscohost.com.proxy.library.nd.edu/login.aspx?direct=true&db=aph&AN=70162084&site=ehost-live}.

\bibitem[Hersbach et~al.(2020)Hersbach, Bell, Berrisford, Hirahara, Horányi,
  Muñoz-Sabater, Nicolas, Peubey, Radu, Schepers, Simmons, Soci, Abdalla,
  Abellan, Balsamo, Bechtold, Biavati, Bidlot, Bonavita, De~Chiara, Dahlgren,
  Dee, Diamantakis, Dragani, Flemming, Forbes, Fuentes, Geer, Haimberger,
  Healy, Hogan, Hólm, Janisková, Keeley, Laloyaux, Lopez, Lupu, Radnoti,
  de~Rosnay, Rozum, Vamborg, Villaume, and Thépaut]{her20}
Hans Hersbach, Bill Bell, Paul Berrisford, Shoji Hirahara, András Horányi,
  Joaquín Muñoz-Sabater, Julien Nicolas, Carole Peubey, Raluca Radu, Dinand
  Schepers, Adrian Simmons, Cornel Soci, Saleh Abdalla, Xavier Abellan,
  Gianpaolo Balsamo, Peter Bechtold, Gionata Biavati, Jean Bidlot, Massimo
  Bonavita, Giovanna De~Chiara, Per Dahlgren, Dick Dee, Michail Diamantakis,
  Rossana Dragani, Johannes Flemming, Richard Forbes, Manuel Fuentes, Alan
  Geer, Leo Haimberger, Sean Healy, Robin~J. Hogan, Elías Hólm, Marta
  Janisková, Sarah Keeley, Patrick Laloyaux, Philippe Lopez, Cristina Lupu,
  Gabor Radnoti, Patricia de~Rosnay, Iryna Rozum, Freja Vamborg, Sebastien
  Villaume, and Jean-Noël Thépaut.
\newblock The era5 global reanalysis.
\newblock \emph{Quarterly Journal of the Royal Meteorological Society},
  146\penalty0 (730):\penalty0 1999--2049, 2020.
\newblock \doi{10.1002/qj.3803}.

\bibitem[Higdon(2002)]{hig02}
Dave Higdon.
\newblock Space and space-time modeling using process convolutions.
\newblock In Clive~W. Anderson, Vic Barnett, Philip~C. Chatwin, and Abdel~H.
  El-Shaarawi, editors, \emph{Quantitative Methods for Current Environmental
  Issues}, pages 37--56, London, 2002. Springer London.

\bibitem[Ho et~al.(2012)Ho, Stephenson, Collins, Ferro, and Brown]{ho12}
Chun~Kit Ho, David~B. Stephenson, Matthew Collins, Christopher A.~T. Ferro, and
  Simon~J. Brown.
\newblock {Calibration strategies: a source of additional uncertainty in
  climate change projections}.
\newblock \emph{Bulletin of the American Meteorological Society}, 93\penalty0
  (1):\penalty0 21--26, 01 2012.
\newblock \doi{10.1175/2011BAMS3110.1}.

\bibitem[{International Renewable Energy Agency}(2018)]{ire18}
{International Renewable Energy Agency}.
\newblock Renewable energy statistics, 2018.
\newblock
  https://irena.org/publications/2018/Jul/Renewable-Energy-Statistics-2018.

\bibitem[{International Renewable Energy Agency}(2019)]{ire19}
{International Renewable Energy Agency}.
\newblock Renewable energy market analysis: Gcc, 2019.
\newblock https://www.irena.org/publications/2019.

\bibitem[IPCC(2014)]{ipc14}
IPCC.
\newblock Part a: Global and sectoral aspects.
\newblock In \emph{AR5 Climate Change 2014: Impacts, Adaptation, and
  Vulnerability}. Cambridge University Press., 2014.

\bibitem[Jeong et~al.(2018)Jeong, Castruccio, Crippa, and Genton]{jeo18}
Jaehong Jeong, Stefano Castruccio, Paola Crippa, and Marc~G. Genton.
\newblock Reducing storage of global wind ensembles with stochastic generators.
\newblock \emph{Annals of Applied Statistics}, 12\penalty0 (1):\penalty0
  490--509, 03 2018.
\newblock \doi{10.1214/17-AOAS1105}.

\bibitem[Jeong et~al.(2019)Jeong, Yan, Castruccio, and Genton]{jeo19}
Jaehong Jeong, Y.~Yan, Stefano Castruccio, and Marc~G. Genton.
\newblock A stochastic generator of global monthly wind energy with {T}ukey
  g-and-h autoregressive processes.
\newblock \emph{Statistica Sinica}, 29:\penalty0 1105--1126, 2019.

\bibitem[Kennedy and O'Hagan(2000)]{ken00}
M.~C. Kennedy and A.~O'Hagan.
\newblock Predicting the output from a complex computer code when fast
  approximations are available.
\newblock \emph{Biometrika}, 87\penalty0 (1):\penalty0 1--13, 2000.

\bibitem[Kennedy and O'Hagan(2001)]{ken01}
Marc~C. Kennedy and Anthony O'Hagan.
\newblock Bayesian calibration of computer models.
\newblock \emph{Journal of the Royal Statistical Society: Series B (Statistical
  Methodology)}, 63\penalty0 (3):\penalty0 425--464, 2001.
\newblock \doi{https://doi.org/10.1111/1467-9868.00294}.

\bibitem[Kim et~al.(2015)Kim, Kwon, and Han]{kim15}
Kue~Bum Kim, Hyun-Han Kwon, and Dawei Han.
\newblock Bias correction methods for regional climate model simulations
  considering the distributional parametric uncertainty underlying the
  observations.
\newblock \emph{Journal of Hydrology}, 530:\penalty0 568--579, 2015.
\newblock ISSN 0022-1694.
\newblock \doi{https://doi.org/10.1016/j.jhydrol.2015.10.015}.
\newblock URL
  \url{http://www.sciencedirect.com/science/article/pii/S002216941500774X}.

\bibitem[Kinley(2017)]{kin17}
Richard Kinley.
\newblock Climate change after paris: from turning point to transformation.
\newblock \emph{Climate Policy}, 17\penalty0 (1):\penalty0 9--15, 2017.
\newblock \doi{10.1080/14693062.2016.1191009}.

\bibitem[Kullback and Leibler(1951)]{kul51}
S.~Kullback and R.~A. Leibler.
\newblock On information and sufficiency.
\newblock \emph{Annals of Mathematical Statistics}, 22\penalty0 (1):\penalty0
  79--86, 03 1951.
\newblock \doi{10.1214/aoms/1177729694}.

\bibitem[Leeds et~al.(2015)Leeds, Moyer, and Stein]{lee15}
W.~B. Leeds, E.~J. Moyer, and M.~L. Stein.
\newblock Simulation of future climate under changing temporal covariance
  structures.
\newblock \emph{Advances in Statistical Climatology, Meteorology and
  Oceanography}, 1\penalty0 (1):\penalty0 1--14, 2015.
\newblock \doi{10.5194/ascmo-1-1-2015}.
\newblock URL \url{https://ascmo.copernicus.org/articles/1/1/2015/}.

\bibitem[Li et~al.(2019)Li, Feng, Xu, Yin, Shi, and Qi]{li}
Delei Li, Jianlong Feng, Zhenhua Xu, Baoshu Yin, Hongyuan Shi, and Jifeng Qi.
\newblock Statistical bias correction for simulated wind speeds over
  cordex-east asia.
\newblock \emph{Earth and Space Science}, 6\penalty0 (2):\penalty0 200--211,
  2019.
\newblock \doi{10.1029/2018EA000493}.
\newblock URL
  \url{https://agupubs.onlinelibrary.wiley.com/doi/abs/10.1029/2018EA000493}.

\bibitem[Mehrotra and Sharma(2016)]{meh16}
Rajeshwar Mehrotra and Ashish Sharma.
\newblock A multivariate quantile-matching bias correction approach with auto-
  and cross-dependence across multiple time scales: Implications for
  downscaling.
\newblock \emph{Journal of Climate}, 29\penalty0 (10):\penalty0 3519--3539,
  2016.

\bibitem[Nguyen et~al.(2019)Nguyen, Mehrotra, and Sharma]{ngu19}
H.~Nguyen, R.~Mehrotra, and A.~Sharma.
\newblock Correcting systematic biases across multiple atmospheric variables in
  the frequency domain.
\newblock \emph{Climate Dynamics}, 52:\penalty0 1283–1298, 2019.

\bibitem[{NREP}(2018)]{nre18}
{NREP}.
\newblock Saudi arabia renewable energy targets and long term visibility,
  national renewable energy program, 2018.

\bibitem[Nurunnabi(2017)]{nur17}
M.~Nurunnabi.
\newblock Transformation from an oil-based economy to a knowledge-based economy
  in saudi arabia: the direction of saudi vision 2030.
\newblock \emph{Journal of the Knowledge Economy}, 8\penalty0 (2):\penalty0
  536--64, 2017.

\bibitem[Paciorek and Schervish(2006)]{pac06}
Christopher~J. Paciorek and Mark~J. Schervish.
\newblock Spatial modelling using a new class of nonstationary covariance
  functions.
\newblock \emph{Environmetrics}, 17\penalty0 (5):\penalty0 483--506, 2006.
\newblock \doi{10.1002/env.785}.

\bibitem[Palacios and Steel(2006)]{pal06}
M.~Blanca Palacios and Mark F.~J Steel.
\newblock Non-gaussian bayesian geostatistical modeling.
\newblock \emph{Journal of the American Statistical Association}, 101\penalty0
  (474):\penalty0 604--618, 2006.
\newblock \doi{10.1198/016214505000001195}.

\bibitem[Poppick et~al.(2016)Poppick, McInerney, Moyer, and Stein]{pop16}
Andrew Poppick, David~J. McInerney, Elisabeth~J. Moyer, and Michael~L. Stein.
\newblock Temperatures in transient climates: Improved methods for simulations
  with evolving temporal covariances.
\newblock \emph{Annals of Applied Statistics}, 10\penalty0 (1):\penalty0
  477--505, 03 2016.
\newblock \doi{10.1214/16-AOAS903}.

\bibitem[Rehman et~al.(2007)Rehman, El-Amin, Ahmad, Shaahid, Al-Shehri, and
  Bakhashwain]{rehman}
S.~Rehman, I.M. El-Amin, F.~Ahmad, S.M. Shaahid, A.M. Al-Shehri, and J.M.
  Bakhashwain.
\newblock Wind power resource assessment for rafha, saudi arabia.
\newblock \emph{Renewable and Sustainable Energy Reviews}, 11\penalty0
  (5):\penalty0 937--950, 2007.
\newblock ISSN 1364-0321.
\newblock \doi{https://doi.org/10.1016/j.rser.2005.07.003}.
\newblock URL
  \url{http://www.sciencedirect.com/science/article/pii/S1364032105000717}.

\bibitem[{REN21 Secretariat}(2018)]{ren18}
{REN21 Secretariat}.
\newblock Renewables 2018 - global status report, 2018.
\newblock Paris, France.

\bibitem[Risser and Calder(2017)]{ris17}
Mark Risser and Catherine Calder.
\newblock Local likelihood estimation for covariance functions with
  spatially-varying parameters: The convospat package for r.
\newblock \emph{Journal of Statistical Software, Articles}, 81\penalty0
  (14):\penalty0 1--32, 2017.
\newblock ISSN 1548-7660.
\newblock \doi{10.18637/jss.v081.i14}.

\bibitem[Stein(1999)]{ste99}
M.L. Stein.
\newblock \emph{Interpolation of Spatial Data: Some Theory for Kriging}.
\newblock Springer, NY, 1999.

\bibitem[Tagle et~al.(2020{\natexlab{a}})Tagle, Genton, Yip, Mostamandi,
  Stenchikov, and Castruccio]{tag21}
F.~Tagle, M.G. Genton, A.~Yip, S.~Mostamandi, G.~Stenchikov, and S.~Castruccio.
\newblock A high-resolution bilevel skew-$t$ stochastic generator for assessing
  saudi arabia's wind energy resources (with discussion).
\newblock \emph{Environmetrics}, 31:\penalty0 e2628, 2020{\natexlab{a}}.
\newblock \doi{10.1002/env.2628}.

\bibitem[Tagle et~al.(2019)Tagle, Castruccio, Crippa, and Genton]{tag19}
Felipe Tagle, Stefano Castruccio, Paola Crippa, and Marc~G. Genton.
\newblock A non-gaussian spatio-temporal model for daily wind speeds based on a
  multi-variate skew-$t$ distribution.
\newblock \emph{Journal of Time Series Analysis}, 40\penalty0 (3):\penalty0
  312--326, 2019.
\newblock \doi{10.1111/jtsa.12437}.

\bibitem[Tagle et~al.(2020{\natexlab{b}})Tagle, Castruccio, and Genton]{tag20b}
Felipe Tagle, Stefano Castruccio, and Marc~G. Genton.
\newblock A hierarchical bi-resolution spatial skew-t model.
\newblock \emph{Spatial Statistics}, 35:\penalty0 100398, 2020{\natexlab{b}}.
\newblock ISSN 2211-6753.
\newblock \doi{https://doi.org/10.1016/j.spasta.2019.100398}.

\bibitem[Taylor et~al.(2012)Taylor, Stouffer, and Meehl]{cmip5}
Karl~E. Taylor, Ronald~J. Stouffer, and Gerald~A. Meehl.
\newblock An overview of cmip5 and the experiment design.
\newblock \emph{Bulletin of the American Meteorological Society}, 93\penalty0
  (4):\penalty0 485--498, 04 2012.
\newblock ISSN 0003-0007.
\newblock \doi{10.1175/BAMS-D-11-00094.1}.
\newblock URL \url{https://doi.org/10.1175/BAMS-D-11-00094.1}.

\bibitem[Teutschbein and Seibert(2012)]{teu}
Claudia Teutschbein and Jan Seibert.
\newblock Bias correction of regional climate model simulations for
  hydrological climate-change impact studies: Review and evaluation of
  different methods.
\newblock \emph{Journal of Hydrology}, 456-457:\penalty0 12--29, 2012.
\newblock ISSN 0022-1694.
\newblock \doi{https://doi.org/10.1016/j.jhydrol.2012.05.052}.
\newblock URL
  \url{http://www.sciencedirect.com/science/article/pii/S0022169412004556}.

\bibitem[Tuo and Wu(2015)]{tuo15}
Rui Tuo and C.~F.~Jeff Wu.
\newblock {Efficient calibration for imperfect computer models}.
\newblock \emph{The Annals of Statistics}, 43\penalty0 (6):\penalty0 2331 --
  2352, 2015.
\newblock \doi{10.1214/15-AOS1314}.

\bibitem[{UNFCCC}(2020)]{unf20}
{UNFCCC}.
\newblock Intended nationally determined contributions: {K}ingdom of {S}audi
  {A}rabia, 2020.
\newblock
  www4.unfccc.int/sites/submissions/INDC/Published\%20Documents/Saudi\%20Arabia/1/KSA-INDCs\%20English.pdf.

\bibitem[van Vuuren et~al.(2011)van Vuuren, Edmonds, Kainuma, Riahi, Thomson,
  Hibbard, Hurtt, Kram, Krey, Lamarque, Masui, Meinshausen, Nakicenovic, Smith,
  and Rose]{van11}
Detlef~P. van Vuuren, Jae Edmonds, Mikiko Kainuma, Keywan Riahi, Allison
  Thomson, Kathy Hibbard, George~C. Hurtt, Tom Kram, Volker Krey, Jean-Francois
  Lamarque, Toshihiko Masui, Malte Meinshausen, Nebojsa Nakicenovic, Steven~J.
  Smith, and Steven~K. Rose.
\newblock {T}he representative concentration pathways: an overview.
\newblock \emph{Climatic Change}, 109:\penalty0 5--31, 2011.

\bibitem[Vassallo et~al.(2020)Vassallo, Krishnamurthy, and Fernando]{vas19}
D.~Vassallo, R.~Krishnamurthy, and H.~J.~S. Fernando.
\newblock Decreasing wind speed extrapolation error via domain-specific feature
  extraction and selection.
\newblock \emph{Wind Energy Science}, 5\penalty0 (3):\penalty0 959--975, 2020.
\newblock \doi{10.5194/wes-5-959-2020}.

\bibitem[{Wang} et~al.(2009){Wang}, {Kulkarni}, and {Verdu}]{wang}
Q.~{Wang}, S.~R. {Kulkarni}, and S.~{Verdu}.
\newblock Divergence estimation for multidimensional densities via $k$-nearest
  neighbor distances.
\newblock \emph{IEEE Transactions on Information Theory}, 55\penalty0
  (5):\penalty0 2392--2405, 2009.

\bibitem[{World Bank}(2020)]{wb20}
{World Bank}.
\newblock Energy use (kg of oil equivalent per capita), 2020.
\newblock https://data.worldbank.org/indicator/EG.USE.PCAP.KG.OE.

\bibitem[Yeo and Johnson(2000)]{yeo00}
In-Kwon Yeo and Richard~A. Johnson.
\newblock A new family of power transformations to improve normality or
  symmetry.
\newblock \emph{Biometrika}, 87\penalty0 (4):\penalty0 954--959, 2000.

\bibitem[Yuan et~al.(2019)Yuan, Thorarinsdottir, Beldring, Wong, Huang, and
  Xu]{yua19}
Qifen Yuan, Thordis~L. Thorarinsdottir, Stein Beldring, Wai~Kwok Wong, Shaochun
  Huang, and Chong-Yu Xu.
\newblock New approach for bias correction and stochastic downscaling of future
  projections for daily mean temperatures to a high-resolution grid.
\newblock \emph{Journal of Applied Meteorology and Climatology}, 58\penalty0
  (12):\penalty0 2617--2632, 12 2019.
\newblock \doi{10.1175/JAMC-D-19-0086.1}.
\newblock URL \url{https://doi.org/10.1175/JAMC-D-19-0086.1}.

\end{thebibliography}

\end{document}